\documentclass{comjnl}

\usepackage{color}
\usepackage{amsmath}
\usepackage{amssymb}
\usepackage{natbib}
\usepackage{amsbsy}
\usepackage{graphicx}
\usepackage{listings}
\usepackage{pdflscape} 
\usepackage[bottom]{footmisc}
\usepackage[mathscr]{euscript}
\graphicspath{{Figures/}} 
\usepackage{hyperref}
\hypersetup{
    colorlinks=true,
    linkcolor=blue,
    filecolor=blue,      
    urlcolor=blue,
    citecolor=blue,
}

\DeclareMathOperator\SF{\text{SF}}

\begin{document}

\title[Is The Starry Night Turbulent?]{\LARGE{Is The Starry Night Turbulent?}}
\author{James R. Beattie$^{1}$ and Neco Kriel$^{2}$ }
\affiliation{$^{1}$Research School of Astronomy and Astrophysics, Australian National University, Canberra, Australia \\
$^{2}$Science and Engineering Faculty, Queensland University of Technology, Brisbane, Australia } 
\shortauthors{J. R. Beattie \& N. Kriel}

\begin{abstract}
Vincent van Gogh's painting, The Starry Night, is an iconic piece of art and cultural history. The painting portrays a night sky full of stars, with eddies (spirals) both large and small. \cite{Kolmogorov1941}'s description of subsonic, incompressible turbulence gives a model for turbulence that involves eddies interacting on many length scales, and so the question has been asked: is The Starry Night turbulent? To answer this question, we calculate the azimuthally averaged power spectrum of a square region ($1165 \times 1165$ pixels) of night sky in The Starry Night. We find a power spectrum, $\mathcal{P}(k)$, where $k$ is the wavevector, that shares the same features as supersonic turbulence. It has a power-law $\mathcal{P}(k) \propto k^{-2.1\pm0.3}$ in the scaling range, $34 \leq k \leq 80$. We identify a driving scale, $k_\text{D} = 3$, dissipation scale, $k_\nu = 220$ and a bottleneck. This leads us to believe that van Gogh's depiction of the starry night closely resembles the turbulence found in real molecular clouds, the birthplace of stars in the Universe.
\end{abstract}

\maketitle

\section{Introduction}\label{sec:intro}
\footnotetext[1]{contact: beattijr@mso.anu.edu.au} \footnotetext[2]{contact: neco.kriel@connect.qut.edu.au}  Recently, parallels have been drawn between the \cite{Kolmogorov1941}'s turbulence (i.e. subsonic, incompressible, high Reynolds number flow) and the swirls and vortices depicted by Vincent Van Gogh, in particular, van Gogh's The Starry Night \citep{Arag2008}. The Starry Night is of particular interest to the astronomer who knows that stars are born inside of turbulent molecular gas clouds \citep{Scalo1997,Ferriere2001,MacLow2004,Kainulainen2009,Arzoumanian2011,Federrath2012,Federrath2013,Andre2014,Federrath2016,Hacar2018,Mocz2018}. If, by coincidence, van Gogh had constructed a turbulent starry night this would be a pleasing coincidence. Previously, \cite{Arag2008} established that there were scaling relations present in the luminosity values of The Starry Night. However, a detailed power spectrum analysis is still lacking, which is essential to establish if The Starry Night is indeed turbulent.

The aim of this study is to perform this analysis by calculating the azimuthally-averaged power spectrum, $\mathcal{P}(k)$, where $k$ is the wavevector, of the night sky in The Starry Night. This will be essential for establishing whether or not there exists  something that resembles an energy cascade, $\mathcal{P}(k) \propto k^{\alpha}$, with a Kolmogorov power-law scaling, $\alpha = -5/3$, a driving, $k_\text{D}$ and a dissipation scale, $k_{\nu}$, \citep{Kolmogorov1941}. All of which are key features that define turbulent flows.  

This article is organised into the following sections: In \S\ref{sec:TurbTheory} we follow \cite{Kolmogorov1941}'s derivation of the scaling laws of turbulent flows, and indicate which quantities we will need to measure in order understand if The Starry Night indeed depicts turbulent behaviour in the night sky. Next, in \S\ref{sec:Data} we discuss the pre-processing methods that we use on The Starry Night. In \S\ref{sec:Methods} we perform a two-dimensional (2D) power spectrum analysis to determine if the eddy structures are isotropic (an assumption of \citealt{Kolmogorov1941}). Following we use the 2D power spectrum to construct the azimuthally-averaged power spectrum, providing us with a means of directly testing if The Starry Night is turbulent in nature (i.e. if a scaling range, driving scale and dissipation scale exist). Finally, in \S\ref{sec:Findings} we summarise our findings.

\begin{figure*}
\centering
\includegraphics[width=\linewidth]{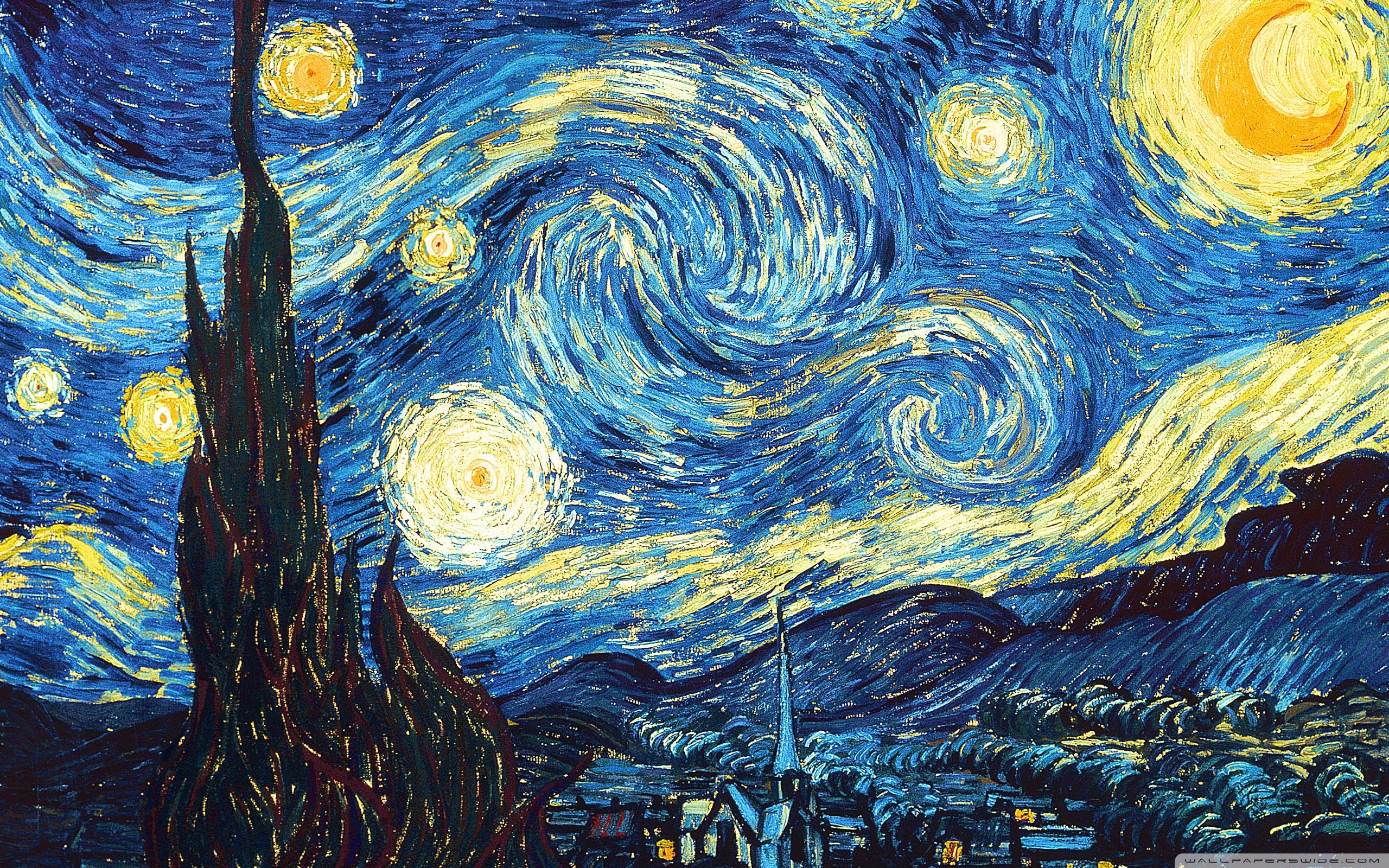}
\caption{ Vincent van Gogh's The Starry Night, accessed from \cite{WallpapersWide2018}. We see eddies painted through the starry night sky that resemble the structures comparable to what we see in turbulent flows.   }
\label{fig:StarryNight}
\end{figure*}

\section{Turbulence Theory}\label{sec:TurbTheory}
In this section we follow \cite{Kolmogorov1941}'s (K41) original arguments to clarify what it means to be turbulent. We first consider eddies (vortices) on length scale $\ell$, with rotation velocity $v_\ell$, which have a turnover time $t_\ell \sim \ell/v_\ell$. We denote the kinetic energy of the eddy at length scale $\ell$ with $\epsilon_\ell$, which means the flow rate of the eddy is approximately,

\begin{equation} \label{eq:turb1}
    \dot{\epsilon_\ell} = \frac{d \epsilon_\ell}{dt} \sim \frac{\epsilon_\ell}{t_\ell} \sim \left( \frac{v_\ell^2}{2} \right) \left( \frac{v_\ell}{\ell} \right) \sim \frac{v_\ell^3}{\ell}.
\end{equation}

If dissipation is negligible, i.e. for high Reynolds number ($Re \sim v/\nu$, where $\nu$ is the dissipation), then $\dot{\epsilon_\ell}$ is conserved and the kinetic energy of the turbulent eddies flows from small to large length scales. This defines the energy cascade (or scaling range, which is more general), 

\begin{equation}
    L > \ell > \ell_\nu,
\end{equation}

where $L$ is the largest scale in the turbulence, and $\ell_\nu$ is the scale which dissipation effects are no longer negligible. This leads to the scaling relation $v \sim (\dot{\epsilon} \ell)^{1/3}$, of the velocity through the energy cascade. We will be searching for such a scale in van Gogh's Starry Night, however, we will not be searching in $\ell$, but rather in $k$, the wavevector space. To construct the above argument in wavevector space we must now introduce the $2^\text{nd}$ order structure function.

The $2^\text{nd}$ order structure function is the average of the square difference in velocities separated by distance $\boldsymbol{\ell}$. It is mathematically defined as, 

\begin{equation} \label{turb2}
   \SF_2(\boldsymbol{\ell}) = \left\langle \right[ \mathbf{v}(\mathbf{r} + \boldsymbol{\ell}) - \mathbf{v}(\mathbf{r}) \left]^2 \right\rangle = \left\langle (\delta\mathbf{v})^2 \right\rangle,
\end{equation}

where the operator $\left\langle \hdots \right\rangle$ indicates the averaging over the ensemble of independent positions $\mathbf{x}$. If the turbulence is statistically isotropic, a question that we will explore in this study, then the averages of $\delta\mathbf{v}^2$ only depend upon the magnitude of the separation distance, $|\boldsymbol{\ell}|$. Clearly $\SF_2(|\boldsymbol{\ell}|) \propto \left\langle v_\ell^2 \right\rangle$ and therefore, from Equation \ref{eq:turb1}, 

\begin{equation}
    \SF_2(|\boldsymbol{\ell}|) \propto (\dot{\epsilon}\ell )^{2/3}.
\end{equation}

The power spectrum, which is further discussed in \S \ref{sec:Methods}, is related to the velocity fluctuations by, 

\begin{equation} 
    \frac{1}{2} \int_0^\infty \left\langle v'_\ell \right\rangle d^3\ell' = \int^0_\infty \mathcal{P}_v (k) d^3 k.  
\end{equation}

where k is the wavevector $\ell = 2\pi/k$. Using the previous proportionality relations and taking the Fourier transform, we find,

\begin{equation}
    \mathcal{P}_v(k) \propto \SF_2 \ell^3 \propto k^{-3}(k^{-1} \dot{\epsilon}^{2/3}) \propto \dot{\epsilon}^{2/3}k^{-11/3}.
\end{equation}

For isotropic turbulence we know that $d^3k \rightarrow 4\pi^2dk$. This means that

\begin{equation}
    \mathcal{P}_v(k) d^3k \propto \mathcal{P}_v(k) k^2 dk  \propto \dot{\epsilon}^{2/3}k^{-5/3} dk
\end{equation}

for power in the modes from $k$ to $k+dk$. The $k^{-5/3}$ scaling is a prediction made by the K41 model. To be clear, we assume that we were looking at scales on $L > \ell > \ell_\nu$, that the flow was incompressible (i.e. no $\rho$ dependence), and that the flow was statistically isotropic. For supersonic spectrum, a similar power-law, $k^{-2}$, was predicted (modelling supersonic turbulence as a series of shocks, which have Fourier transforms $\sim 1/k$) and tested with confirmation in high resolution simulations \citep{Burgers1948,Kritsuk2007,Konstandin2012,Federrath2013,Federrath2018}. The aim of this study will be to identify the scaling range (if one exists) and calculate the power-law $\mathcal{P}(k) \propto k^{\alpha}$.

\begin{figure*}
\centering
\includegraphics[width=\linewidth]{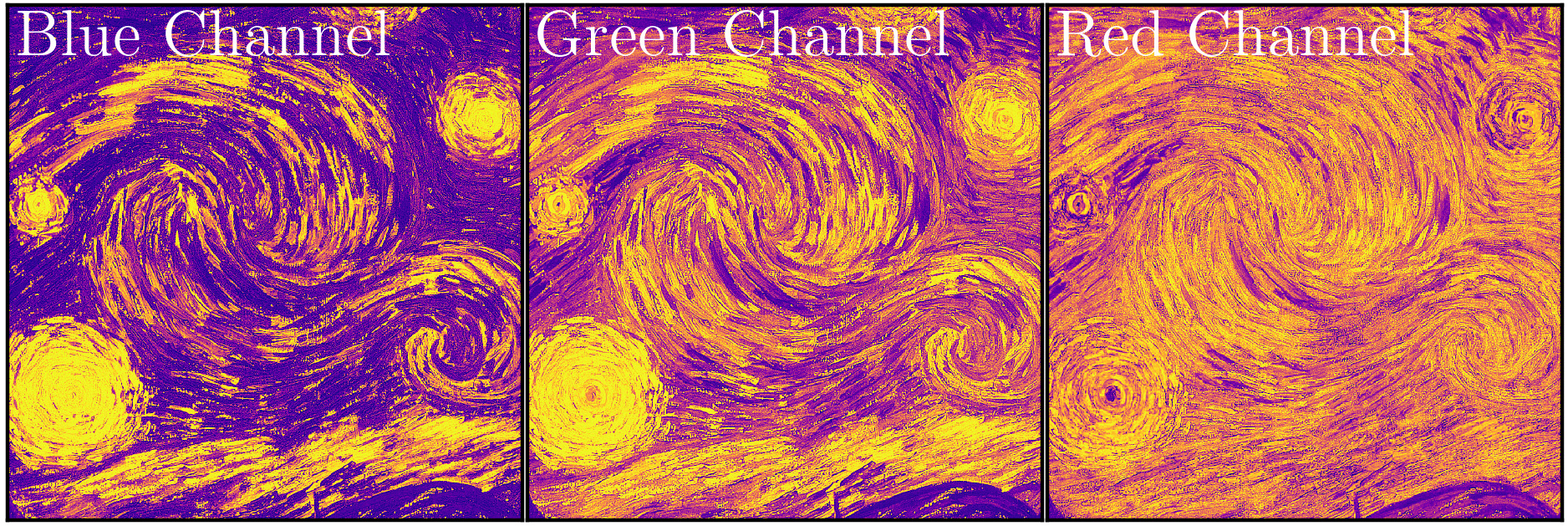}
\caption{ The square $(1165 \times 1165)$ region, separated by channel, that we selected from The Starry Night to encompass the largest eddy structures, whilst omitting the city and the towering structure in the left of Figure \ref{fig:StarryNight} }
\label{fig:Channels}
\end{figure*}

\section{Starry-Night Data} \label{sec:Data}
To analyse the vortices in Van Gogh's Starry Night we first acquire a high resolution digital copy of the painting. We access a $2560 \times 1600$ pixel image from \cite{WallpapersWide2018}, shown in Figure \ref{fig:StarryNight}. There are more structures (i.e. buildings and trees) in the painting than just (possibly) turbulent vortices. For this reason, we focus on a region on the painting that contains the night sky. This limits us to the analysis of the largest square region in the night sky, containing the maximum amount of vortice information. We choose a square region $(1165 \times 1165)$ because then constructing the power spectrum becomes an easier task. The region we select is shown in Figure \ref{fig:Channels}. All three of the channels contain structural information about the nature of the vortices, as shown in Figure \ref{fig:Channels}, so we perform our 2D power spectrum analysis on each of them. In previous studies, averaging is performed at this stage in the processing across the channels, however we restrain ourselves from averaging until we know that there is not a significant difference between them. From herein we will refer to the square region of The Starry Night as $\mathscr{SN}(\boldsymbol{\ell})$, where $\boldsymbol{\ell} \in \mathbb{R}^2$ is the coordinate vector in real-space. Also note whilst K41's model is based upon the power spectrum of the velocity, we can only access pixel values, similar to the analysis in \cite{Arag2008}, hence $\mathscr{SN}(\boldsymbol{\ell})$ is the pixel value at pixel position $\boldsymbol{\ell}$.

\begin{figure*}
\centering
\includegraphics[width=\linewidth]{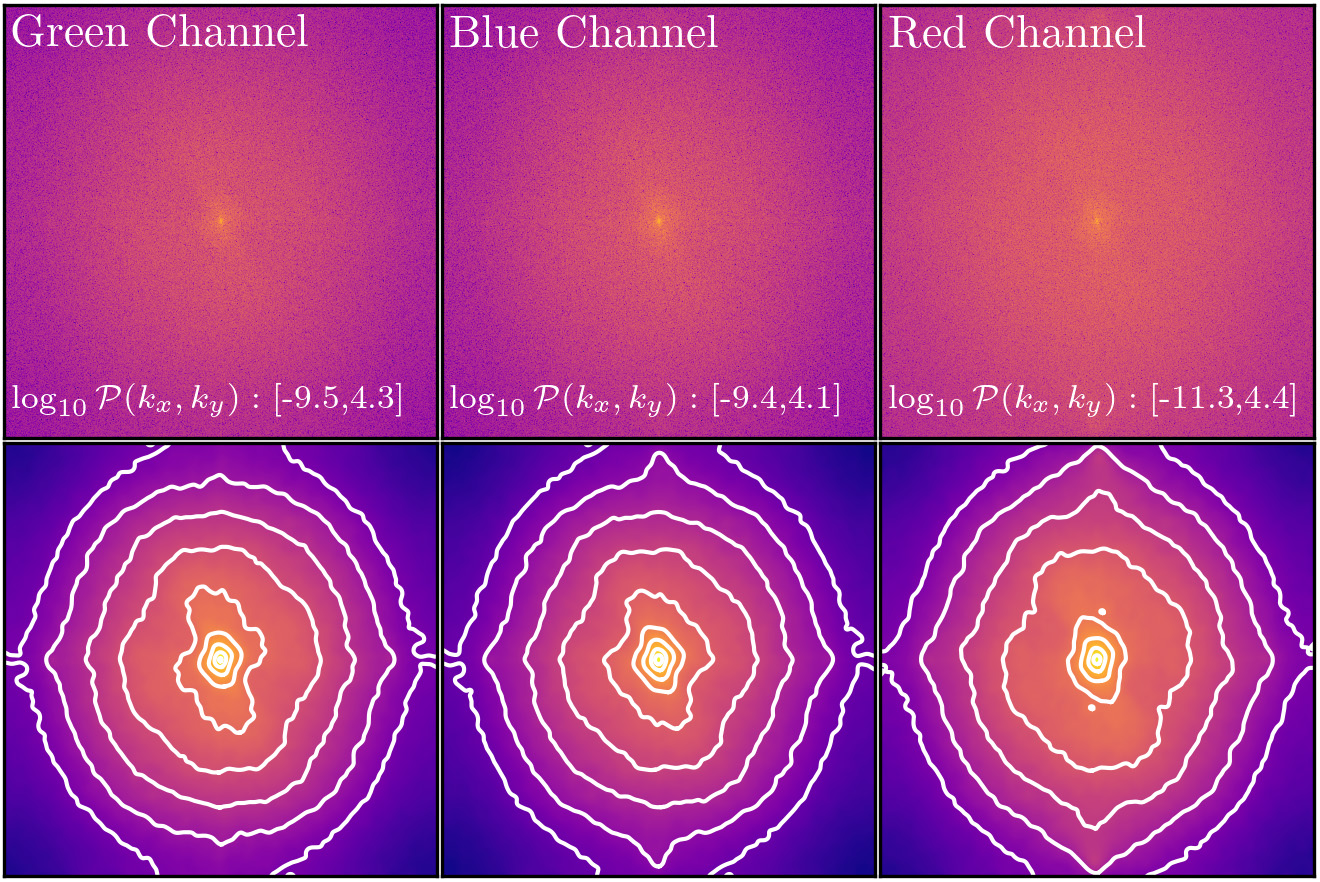}
\caption{ Top: the centred, 2D power spectrum, $\mathcal{P}(\mathbf{k})$ for each of the channels. Bottom: the 2D power spectrum convolved with a Gaussian kernel to smooth the spectrum, and then shown with isobars every $\approx$ 0.5 $\log_{10} \mathcal{P}(\mathbf{k})$. We see that the power spectrum is ansiotropic for small $k$, but becomes more isotropic as $k \gg 1$.}
\label{fig:2DPowerSpec}
\end{figure*}

\section{Power Spectrum Methods}\label{sec:Methods}
\subsection{The 2D Fourier Transform and Power Spectrum} \label{sec:PowerSpec}
To access the scaling relations discussed in \S\ref{sec:TurbTheory} we construct the 2D power spectrum for the vortices shown in Figure \ref{fig:Channels}.

First we calculate the 2D Fourier transform,

\begin{equation} \label{eq:2DFourier}
    \widehat{\mathscr{SN}}(\mathbf{k}) = \frac{1}{n^2} \int \mathscr{SN}(\boldsymbol{\ell}) e^{-i\mathbf{k}\cdot\mathbf{\ell}} d^2 \boldsymbol{\ell},
\end{equation}

where $n = 1165$ is the box size of the region in The Starry Night. The power spectrum is the square of the Fourier transform, analogous to the regular definition of power, which is the square of intensity of a signal. It is therefore given as the square of the intensities of frequencies in the 2D Fourier space, hence we calculate it by,

\begin{equation} \label{eq:powerspec}
    \mathcal{P}(\mathbf{k}) =\widehat{\mathscr{SN}}(\mathbf{k}) \widehat{\mathscr{SN}}^*(\mathbf{k}),
\end{equation}

where $\widehat{\mathscr{SN}}^*(\mathbf{k})$ is the complex conjugate of the Fourier transform. In this study we encode the normalisation of power spectrum into the coefficient of the Fourier transform, shown in equation \ref{eq:2DFourier}. For this reason, equation \ref{eq:2DFourier} deviates slightly from the regular definition of the 2D Fourier transform. Finally, we centre the power spectrum by shifting the $\mathbf{k} = (0,0)$ to the centre of the 2D data. 

The overall goal here is to reduce the 2D power spectrum into a one-dimensional (1D) power spectrum by performing azimuthal averaging, so that we can identify any features that resemble turbulent dynamics. We can only justify averaging over the azimuthal angle for particular $|\mathbf{k}|$ if the spectrum is isotropic, i.e. invariant under rotations about the origin. To test this condition, we first apply a Gaussian smoothing kernel (with standard deviation $\sigma = 10$) to the 2D power spectra, to reduce the Gaussian noise. Once we have smoothed the spectrum, we perform a marching squares algorithm on the spectrum to reveal the isobars i.e. the curves of equal power. The smoothing allows us to just focus on the overall shape of the isobars and gives us a qualitative means of observing if the spectra is isotropic. 

\subsection{2D Power Spectrum Results}

In the top three panels of Figure \ref{fig:2DPowerSpec} we plot the 2D power spectra for each of the channels shown in figure \ref{fig:Channels}. In the bottom three panels we shown the smoothed power spectra, with isobars identified using the marching squares algorithm. For large $\mathbf{k}$, and small $\boldsymbol{\ell}$, we see a relatively isotropic power spectrum. Perhaps this is because it is easiest to paint symmetric, small structures, since these can be performed in a single brush stroke. For small $\mathbf{k}$ and hence large $\boldsymbol{\ell}$ (towards the origin of the bottom three panels in Figure \ref{fig:2DPowerSpec}) we see the power spectrum becomes stretched in the $k_y$ direction. This suggests that van Gogh has a preferential orientation in the way he painted large structures in his starry night, possibly from a particular technique that he used to paint it. 

Figure \ref{fig:2DPowerSpec} also suggests that the power spectra from each channel are relatively similar in structure and magnitude, especially on small scales. This is good, because we can average over the power spectra for each channel to produce a single 2D power spectrum, which in turn can be used to create a single azimuthally-averaged power spectrum for The Starry Night, which we do next.

\subsection{The Azimuthally-Averaged Power Spectrum}
Now that we have observed that the 2D power spectrum is similar across each channel we average across them, creating a single 2D power spectrum for the night sky in The Starry Night. We now use this single power spectrum to construct the azimuthal average.

We construct concentric circles, each of radius $|\mathbf{k}| = k$, separated by a single wavevector magnitude, from $k = 0$ to 821. We average over all powers at each of the fixed $k$, through a $\theta \in [0,2\pi]$ rotation in the 2D $k$-space, producing a single power value for each $k$, 
\begin{equation} \label{eq:powerspecav1}
    \Big\langle \mathcal{P}(k)  \Big\rangle_{\theta} = \Big\langle \widehat{\mathscr{SN}}(\mathbf{k}) \widehat{\mathscr{SN}}^*(\mathbf{k}) 2\pi k \Big\rangle.
\end{equation}
Where the operator $\left\langle \hdots \right\rangle_\theta$ is the azimuthal-average around the $k$-space circle.

We also construct the variance,
\begin{equation} \label{eq:powerspecav2}
    \sigma^2 = \Big\langle \mathcal{P}^2(k) \Big\rangle_{\theta} - \Big\langle \mathcal{P}(k)  \Big\rangle_{\theta}^2,
\end{equation}
which we use to construct the $1\sigma$ uncertainties in the power spectrum.

\begin{figure*}
\centering
\includegraphics[width=\linewidth]{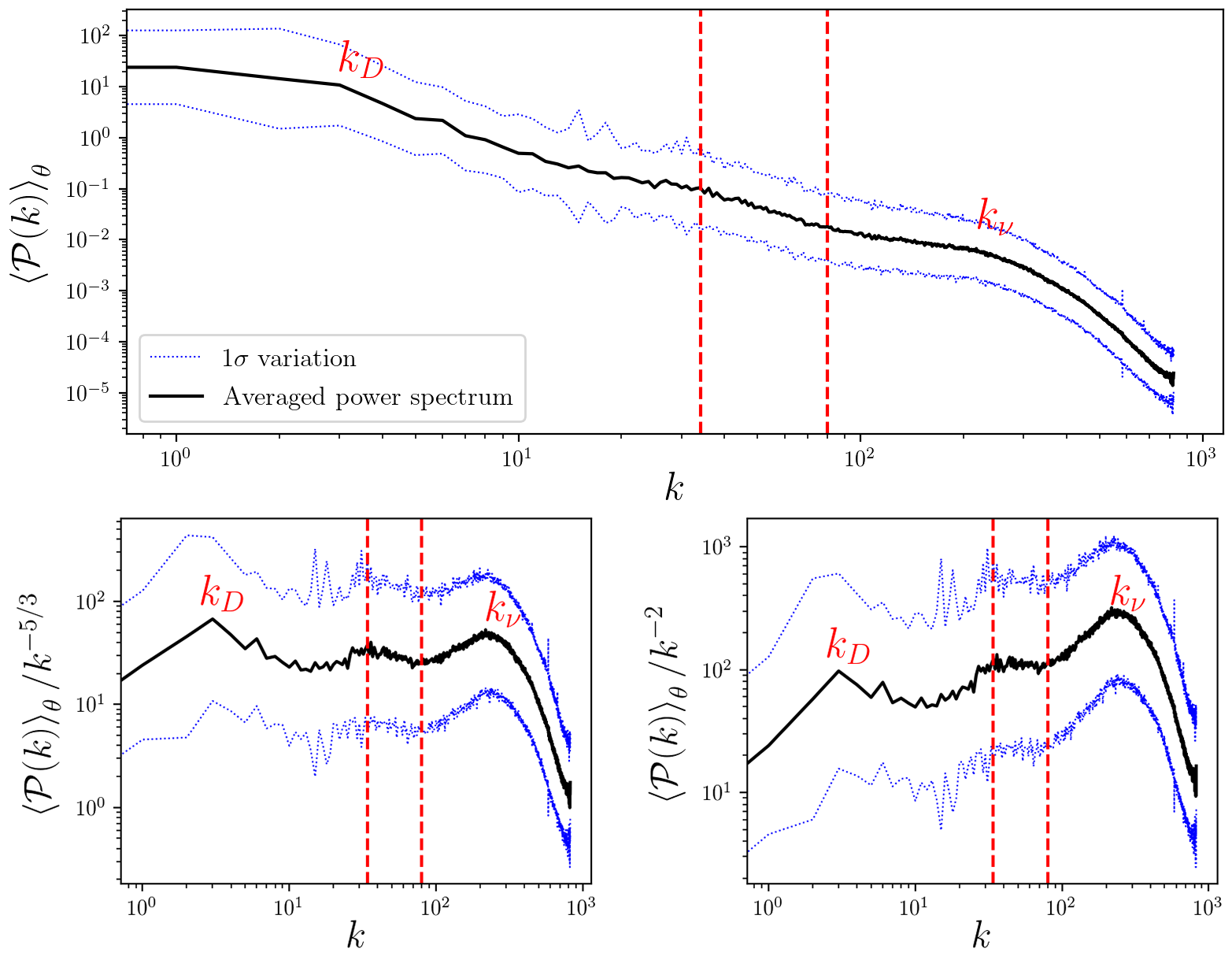}
\caption{\textbf{Top:} The azimuthally averaged power spectrum of van Gogh's Starry Night (shown in black) with 1$\sigma$ uncertainties constructed from the variance of the averaging for each $|\mathbf{k}| = k$ (shown in blue). The power spectrum has similar characteristics to what one would expect from a power spectrum of a field undergoing turbulent dynamics. A driving scale, $k_\text{D}$, and a dissipation scale, $k_\nu$, can be identified. We can also identify a region in the $k$-space that seems to follow a power-law, resembling a scaling range where $\mathcal{P}(k) \sim k^{-2.1 \pm 0.3}$, indicated by the red dashed lines. \textbf{Bottom:} The left plot is scaled by $k^{-5/3}$, to see if the scaling range becomes stationary under a K41 power-law, and the right plot is scaled by $k^{-2}$, to see if the scaling range becomes stationary under Burgers' power-law. We find a small negative gradient under the K41 scaling, but complete stationarity under the \cite{Burgers1948} scaling.}
\label{fig:azipowerspec}
\end{figure*}

\subsection{Azimuthally-Averaged Power Spectrum Results}
In Figure \ref{fig:azipowerspec} we plot the azimuthally-averaged power spectrum from the 3-channel average of the night sky in The Starry Night. We see a peak in the spectrum at small $k$, which we identify as a driving scale $k_\text{D} = 3$. Moving to larger $k$ we see a dip, and then what might be a power-law between $k = 34$ to $k = 80$. We identify this as a scaling-range, noting that it seems to follow a power-law under the log-log transform, as one expects from both K41 and \cite{Burgers1948}. At still larger $k$ we find a dissipation scale $k_\nu = 220$, where the azimuthally-averaged power spectrum quickly tapers off. Before the dissipation scale, more prominently shown the two bottom plots in Figure \ref{fig:azipowerspec}, we also see a build-up of energy at approximately $k_\nu$. This is known as the ``bottle-neck effect", and is realised in real turbulence \citep{Dobler2003}. 

We plot the averaged power spectra with $k^{-5/3}$ and $k^{-2}$ compensations in two bottom plots in Figure \ref{fig:azipowerspec}. Identifying the scaling range, we fit the function $\log \mathcal{P}(k) = \alpha \log k + \beta$, and find that $\alpha = 2.1\pm0.3$. This is why the bottom left plot (K41 scaling) does not quite become stationary, showing a small, negative slope, whereas in the bottom right plot (Burgers' scaling) shows a close to zero slope. $\alpha = 2.1\pm0.3$ is is also consistent with the results from \cite{Arag2008}, who finds that $\SF_2(\ell) \propto \ell^p$, where $p \approx 2$ (\citealt{Arag2008} uses $R$ instead of $\ell$). This suggests that the night sky from The Starry Night is a coincidental depiction of supersonic turbulence ($\mathcal{P}(k) \propto k^{-2}$), a key feature of real molecular clouds in the Universe, which are the birth places of stars \citep{Burgers1948,Kritsuk2007,Konstandin2012,Federrath2012,Federrath2013,Andre2014,Federrath2016,Hacar2018,Mocz2018}.

\section{Summary and Key Findings}\label{sec:Findings}
In this study we determined whether or not the night sky in van Gogh's Starry Night has a power spectrum that resembles a supersonic turbulent flow. First we select a square region of the night sky from The Starry Night. Next, we calculate the two-dimensional (2D) power spectrum for each of the three red-green-blue (RGB) channels. We find the channels have similar power spectra and average over the three spectra to create a single spectrum. We then construct the azimuthally-averaged power spectrum by averaging over concentric circles of radii $|\mathbf{k}| = k$, where $k \in \mathbb{Z}$ is the wavevector. We observe a driving scale, $k_\text{D}$, dissipation scale, $k_\nu$, and scaling range that follows a power-law similar to that of supersonic turbulence. This shows that van Gogh's The Starry Night does exhibit some similarities to turbulence, which happens to be responsible for the real, observable, starry night sky. We summarise the key findings below: \\
\begin{itemize}
    \item The 2D power spectrum of the night sky in The Starry Night is mostly invariant to the RGB channel. It is isotropic for large $\mathbf{k}$, suggesting that van Gogh's night-sky has small, symmetrical structures. For small $\mathbf{k}$, corresponding to large scales in the painting, the power spectrum is squeezed in the $k_y$ direction. This may be due to a particular technique that van Gogh used to paint the large eddies in the night-sky. \\[1em]
    \item We identify characteristic scales in the azimuthally-averaged power spectrum that may resemble a driving scale at $k = 3$, a dissipation scale at $k = 220$ and scaling range between $34 \leq k \leq 80$. We also identify a ``bottleneck-effect" at approximately $k_\nu$. These realisations support the case that The Starry Night is a depiction of a turbulent flow. \\[1em]
    \item The slope of the power spectrum in the identified scaling range is $-2.1 \pm 0.3$, similar to the scaling expected in supersonic turbulence, $-2$. The value of the slope agrees with the second order structure function length scaling found in \cite{Arag2008}, suggesting that van Gogh's depiction of the starry night sky is in fact an accurate portrayal of a supersonic, turbulent medium, bursting with stars. 
\end{itemize}

\noindent The code for this study can be found at the GitHub repository: \\
\url{https://github.com/AstroJames/VanGoghsStarryNight.git}

\section*{Acknowledgements}\label{section:acknowledgments}
J.~R.~B. would like to acknowledge the useful discussions he had with Dean Muir, David Galea and, the brief discussion of the power spectrum results with Christoph Federrath.

\bibliographystyle{compj.bst}
\bibliography{references.bib}

\end{document}